\documentclass[12pt]{article}
\usepackage{amssymb,amsmath,fullpage}
\usepackage[cp1251]{inputenc}
\usepackage{citehack}
\usepackage[active]{srcltx}
\usepackage[dvips]{graphicx}

\begin{document}

\begin{center}
\textbf{Exact solutions of the problem of dynamics of a fluid with a free surface located between two approaching vertical walls}

\vspace{1mm}

\textit{Evgenii A. Karabut$^{1,2}$, Elena N. Zhuravleva$^{1,2}$,  Nikolay M. Zubarev$^{3,4}$, Olga V. Zubareva$^{4}$}

\end{center}

\vspace{3mm}

\footnotetext[{1}]{Lavrentyev Institute of Hydrodynamics, Siberian Branch, Russian Academy of Sciences, Novosibirsk, 630090, Russia}
\footnotetext[{2}]{Novosibirsk State University, Novosibirsk, 630090, Russia}
\footnotetext[{3}]{Lebedev Physical Institute, Russian Academy of Sciences, Moscow, 119991, Russia}
\footnotetext[{4}]{Institute of Electrophysics, Ural Branch, Russian Academy of Sciences, Ekaterinburg, 620016, Russia }

\begin{abstract}
Exact solutions of a classical problem of a plane unsteady potential flow of an ideal incompressible fluid with a free boundary are presented. The fluid occupies a semi-infinite strip bounded by the free surface (from above) and (from the sides) by two solid vertical walls approaching each other with a constant velocity. The solutions are obtained for a situation where the capillary and gravity forces are absent, and the fluid motion is completely determined by the motion of the walls. The solutions contain an arbitrary function, which allows one to describe the nonlinear evolution of perturbations of an arbitrary shape for an initially flat horizontal surface of the fluid. Examples of exact solutions corresponding to the formation of bubbles, cuspidal points, and droplets are considered.
\end{abstract}


\section{Introduction}
Despite an almost two-hundred-year history of the problem of a plane unsteady potential flow of an ideal incompressible fluid with a free boundary (Stokes 1847), only a few exact particular solutions are known. The most well-known class of exact unsteady solutions, flows with a linear field of velocities, was discovered by Dirichlet (1861). A classification of such solutions for the two-dimensional case was performed by Ovsiannikov (1967) and Longuet-Higgins (1972). The free boundary for the corresponding solutions is a second-order curve: ellipse, hyperbola, or straight line (in the degenerate case). In our previous studies (Karabut \& Zhuravleva 2014, Zubarev \& Karabut 2018), we found a new class of unsteady flows with a nonlinear field of velocities, which allows the above-mentioned constraint on the free surface geometry to be eliminated. A similar class of solutions was independently discovered by Zakharov (2020). It should be noted that a significant drawback of these solutions is their ``exotic" nature, as was named by Zakharov (2020): the fluid velocity in flows described by these solutions tends to infinity at the periphery. As a consequence, the resultant flows are mainly of academic rather than of practical interest.

It is demonstrated in the present work that the issue of conditions at infinity can be resolved by considering combined boundary conditions, where, in contrast to (Karabut \& Zhuravleva 2014, Zubarev \& Karabut 2018), the fluid boundary is not completely free. The development of the approaches proposed in recent publications (Zhuravleva \textit{et al.} 2019, Karabut \textit{et al.} 2020) made it possible to find exact nontrivial solutions of the problem of a plane unsteady potential flow of an ideal incompressible fluid with a free boundary, which is located between two approaching impermeable solid vertical walls. The fluid motion is induced both by inertia and by the action of moving solid walls. A noticeable specific feature of the resultant flows is a possibility of their description by the Hopf equation for the complex velocity of the fluid. Owing to its integrability, it is fairly easy to study singular points of the solutions (Caflisch \textit{et al.} 1993,  Kuznetsov \textit{et al.} 1993,  Zubarev \& Kuznetsov 2014). These singularities migrate with time and can finally reach the free surface, thus, leading to solution destruction (Tanveer 1991, Lushnikov \& Zubarev 2018, Dyachenko \textit{et al.} 2021) (see also Bensimon (1986), Howison (1986) for Hele-Shaw flows). Surface deformations induced by singularities are typical, e.g., for Stokes waves with an almost maximum amplitude (Lushnikov 2016). Singularities on the free boundary of breaking standing waves were considered by Baker-Xie (2011). It was demonstrated  by Karabut \textit{et al.} (2019) that singularities reaching the free boundary can initiate the formation of cumulative jets.

\section{Basic equations}

For plane flows of an ideal incompressible fluid, the continuity equation and the potentiality condition are written as
\begin{equation} \label{Eq_1}
\frac{\partial u}{\partial x} +\frac{\partial v}{\partial y} =0,
\quad \quad
\frac{\partial v}{\partial x} -\frac{\partial u}{\partial y} =0.
\end{equation}
Here $u(x,y,t)$ and $v(x,y,t)$ are the $x$ and $y$ components of the velocity vector of the fluid, respectively; $x$, $y$, and $t$ are the Cartesian coordinates and time. Equations \eqref{Eq_1} are the Cauchy-Riemann relations for the function
$$
U(z,t)=u(x,y,t)- \textrm{i} v(x,y,t).
$$
This means that the function $U(z,t)$, which is called the complex velocity, is an analytical function of the complex variable $z=x+ \textrm{i} y$ in the domain occupied by the fluid.

The fluid dynamics in the absence of external forces is determined by the Euler equations
\begin{equation} \label{Eq_3}
\frac{\partial u}{\partial t} +u\,\frac{\partial u}{\partial x}
+v\,\frac{\partial u}{\partial y}
=-\frac{1}{\rho } \frac{\partial p}{\partial x}, \qquad
\frac{\partial v}{\partial t} +u\,\frac{\partial v}{\partial x}
+v\,\frac{\partial v}{\partial y}
=-\frac{1}{\rho } \frac{\partial p}{\partial y},
\end{equation}
where $p(x,y,t)$ is the pressure, and $\rho ={\rm const}$ is the fluid density. Equations \eqref{Eq_1}--\eqref{Eq_3} should be solved together with the boundary conditions. The fluid if bounded at the top by the free boundary, which consists of the same fluid particles for all $t$. If the free boundary is defined by the equation $h(x,y,t)=0,$ then the following (kinematic) condition should be valid:
\begin{equation} \label{Eq_4}
\frac{{\rm d}h}{{\rm d}t} \equiv \frac{\partial h}{\partial t}
+u\,\frac{\partial h}{\partial x} +v\,\frac{\partial h}{\partial y} =0
\; \quad {\rm at}\quad \;
h(x,y,t)=0
\end{equation}
(${\rm d}/{\rm d}t$ means the total derivative). In addition to the kinematic condition, the free surface should satisfy the dynamic condition. We consider a situation where the fluid motion is induced by the motion of the side walls bounding the fluid, whereas the gravity and capillary forces are absent. In this situation, the dynamic condition reduces to the condition of a constant pressure:
\begin{equation} \label{Eq_5}
p={\rm const}\; \quad {\rm at}\quad \; h(x,y,t)=0.
\end{equation}
Let us denote the absolute value of the velocity of each wall by $V$ ($V>0$). The positions of the walls are determined by the equations $x=\pm Vt$, i.e., they collide at the time instant $t=0$ (we are interested in the fluid behavior at $t<0$). The kinematic conditions on the walls are $u|_{x=\pm Vt} =\pm V$ or, in the complex form,
\begin{equation} \label{Eq_6}
\mbox{Re}\,U=\pm V\; \quad {\rm at}\quad \; \mbox{Re}\,z=\pm Vt.
\end{equation}
Finally, we require that $v\to -y/t$ at an infinite depth, $y\to -\infty $, i.e., the fluid is squeezed downward as the walls approach each other.

The problem formulated above admits an exact solution for which the free surface of the fluid is flat. The fluid occupies a semi-infinite strip $y<0$ and $Vt<x<-Vt$. The segment $y=0$ and $Vt<x<-Vt$ is the free boundary. As $t$ increases (let us recall that $t<0$), the strip width decreases and finally vanishes at $t=0$, when the walls collide. The complex velocity is defined by the self-similar relation
\begin{equation} \label{Eq_7}
U=z/t.
\end{equation}
In this case, the pressure and velocity components are described by the equations
$p=-\rho y^{2} /t^{2},$
$ u=x/t,$
$v=-y/t,$
which ensures satisfaction of \eqref{Eq_1}--\eqref{Eq_6}, as it can be easily seen. Clearly, the velocity field acquires a singular character by the time $t=0$ owing to the collision of the walls.

The unsteady flow with a plane free boundary and a linear velocity field described by \eqref{Eq_7} was studied earlier by Ovsiannikov (1967) and  Longuet-Higgins (1972). One can easily find the stream function and demonstrate that the streamlines are hyperbolas. A flow of this type is shown in Fig.~1.

\begin{figure}
\begin{center}
\resizebox{0.5\textwidth}{!}{\includegraphics{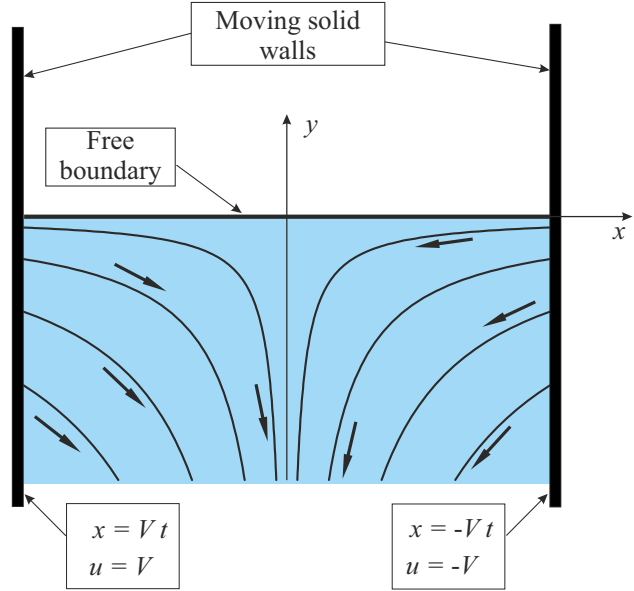}}
\end{center}
\caption{Flow \eqref{Eq_7} at a certain time instant $t<0$. The streamlines and the velocity vector directions are shown schematically. The halftone area means the domain occupied by the fluid. The flow is initiated by horizontal motion of two solid walls with a velocity $V$.}
\label{base}
\end{figure}

\section{Exact solutions}

It was demonstrated (Zubarev \& Karabut 2018 , Zhuravleva \textit{et al.} 2019, Karabut \textit{et al.} 2020) that certain progress in the description of plane flows with a free boundary can be provided by using the hodograph transformation, i.e., choosing  the variable $U$ as an independent variable and the variable $z$ as an unknown function. This transformation is convenient for the problem considered in the present study because the moving boundaries $x=\pm Vt$ of the domain occupied by the fluid in the plane $z$ become fixed after the transition to the hodograph plane $U$: $u=\pm V$.

After the hodograph transformation, the exact particular solution \eqref{Eq_7} obviously takes the form
\begin{equation} \label{Eq_8}
z=Ut.
\end{equation}
This expression determines the undisturbed (base) solution of the problem. Clearly, the disturbed flow can be presented in the general case as
\begin{equation} \label{Eq_8dop}
z=Ut+f(U,t),
\end{equation}
where $f$ is a certain function determining the deviation of the flow from the self-similar one \eqref{Eq_8}. It is demonstrated in the present paper that a wide class of exact solutions of the problem can be found by confining the study to a situation where $\partial f/\partial t=0$, i.e., the disturbance $f$ does not explicitly depend on time.

Thus, we seek the solution for the disturbed flow in the form
\begin{equation} \label{Eq_9}
z=Z(U,t)=Ut+F(U),
\end{equation}
where $F$ is an unknown function of the complex velocity $U$. We do not require that the disturbance $F$ should be small, i.e., we consider the evolution of disturbances of arbitrary amplitudes and shaped.

Let us check whether presentation \eqref{Eq_9} can be consistent with the initial equations of motion \eqref{Eq_1}--\eqref{Eq_6}. It is immediately seen from \eqref{Eq_6} that the following condition should be satisfied on two straight lines $u=\pm V$ in the hodograph plane $U$:
\begin{equation} \label{Eq_10}
\mbox{Re}\, F(U)|_{u=\pm V} =0.
\end{equation}
If we choose the function $F$ satisfying condition \eqref{Eq_10}, the kinematic conditions on the moving walls $x=\pm Vt$ will be automatically fulfilled.

Now let us discuss the conditions on the free surface of the fluid. We rewrite \eqref{Eq_9} via the inverse function $G=F^{-1}$:
\begin{equation} \label{Eq_11}
U(z,t)=G(z-Ut).
\end{equation}
One can easily see that the implicit expression \eqref{Eq_11} is the known solution of the Hopf equation
\begin{equation} \label{Eq_12}
\frac{\partial U}{\partial t} +U\,\frac{\partial U}{\partial z} =0,
\end{equation}
obtained by the method of characteristics. Let us demonstrate that the flow defined by \eqref{Eq_12} or \eqref{Eq_9} ensures that both the kinematic \eqref{Eq_4} and dynamic \eqref{Eq_5} conditions are satisfied.

The Euler equations \eqref{Eq_3} can be written in the complex form as
\begin{equation} \label{Eq_13}
\frac{{\rm d}U}{{\rm d}t}
\equiv \frac{\partial U}{\partial t}
+\overline{U}\,\frac{\partial U}{\partial z}
=-\frac{1}{\rho } \left(\frac{\partial p}{\partial x}
- \textrm{i} \frac{\partial p}{\partial y} \right),
\end{equation}
where the bar means complex conjugation. Subtraction of \eqref{Eq_13} and \eqref{Eq_12} yields
\[
2 \textrm{i} v\,\frac{\partial U}{\partial z}
\equiv -2 \textrm{i} \,\mbox{Im}\,U\, \frac{\partial U}{\partial z}
=-\frac{1}{\rho } \left(\frac{\partial p}{\partial x}
- \textrm{i} \frac{\partial p}{\partial y} \right).
\]
It is clearly seen from here that the condition $\nabla p=0$ ($\nabla \equiv \left(\partial /\partial x,\; \partial /\partial y\right)$) on the surface $\mbox{Im}\,U\equiv -v=0$ is satisfied for the flow described by the Hopf equation \eqref{Eq_12}. Then $p={\rm const}$ on this surface, and the dynamic boundary condition \eqref{Eq_5} is automatically satisfied. The validity of the kinematic condition \eqref{Eq_4} immediately follows from \eqref{Eq_13}, where ${\rm d}U/{\rm d}t=0$ for the zero right-hand side, i.e., for the flow considered here, the fluid particles on the surface $\mbox{Im}\,U\equiv -v=0$ move without acceleration (with a constant velocity). In particular, ${\rm d}v/{\rm d}t=0$, which is actually the kinematic boundary condition in choosing the function $h\equiv v$. Thus, the condition $v=0$ for the flow described by  \eqref{Eq_9} defines the free surface in the hodograph plane (see also Zhuravleva \textit{et al.} (2019), Karabut \textit{et al.} (2020)).

It is essential for the further analysis of the flows describes by  \eqref{Eq_9} with condition \eqref{Eq_10} that the domain occupied by the fluid is unchanged in the hodograph plane $U$. In this plane, the fluid occupies a semi-infinite strip:
\begin{equation} \label{Eq_14}
v<0,\quad \quad -V<u<V.
\end{equation}
It is owing to this property of the unsteady flows under consideration that their description with the use of the hodograph transformation is extremely convenient.

Finally, let us discuss which constraints on the function $F$ are imposed by the requirement that the complex velocity $U$ should be a holomorphic function of the variable $z$ in the domain occupied by the fluid (the analyticity of the function $U$ follows from conditions \eqref{Eq_1}). It follows from the properties of inverse functions that the function $U$ has no singular points in the flow domain \eqref{Eq_14} if the condition $\partial Z/\partial U\ne 0$ is satisfied. In terms of the function $F$, this means that the function $F$ should be analytical and
\begin{equation} \label{Eq_15}
\frac{\partial F(U)}{\partial U} \ne -t.
\end{equation}
In turn, this means that the singular points defined by the condition $\partial F/\partial U=-t$ should be located outside the flow domain.

Condition \eqref{Eq_15} is easily satisfied. If the derivative $\partial F/\partial U$ is finite in the domain described by \eqref{Eq_14} (it is of principal importance here that the domain occupied by the fluid is fixed in the hodograph plane $U$), then condition \eqref{Eq_15} is automatically satisfied for sufficiently large values of $\left|t\right|$. Indeed, let the function $F$ be chosen in such a way that the following condition is valid in the half-strip \eqref{Eq_14}:
\[
\mbox{Re}\left(\frac{\partial F}{\partial U} \right)\le T
\]
($T$ is a certain positive value). Then the condition $\partial F/\partial U=-t$ defining singular points cannot be satisfied in the flow domain at $t<-T$ in principle. Condition \eqref{Eq_15} can be violated at small values of $\left|t\right|$ (at $-T<t<0$), i.e., when the walls are already fairly close to each other. When the singular point reaches the boundary of the domain described by \eqref{Eq_14}, the solution is destroyed, which may be accompanied by the formation of singular points on the free surface of the fluid.

Thus, we demonstrated the following: if we take an analytical function $F$ satisfying conditions \eqref{Eq_10} and \eqref{Eq_15}, which can be easily satisfied, then the implicit expression \eqref{Eq_9} yields the exact solution of problem \eqref{Eq_1}--\eqref{Eq_6} under consideration. This solution can be interpreted as an expression that describes the evolution of disturbances of the base flow \eqref{Eq_7}. It is principally important that we did not require that the disturbances (in our case, the function $F$) should be small, as it is traditionally done in the flow stability analysis. For the solutions derived here, the disturbance amplitude is arbitrary; as a consequence, they describe the nonlinear evolution of disturbances up until the formation of various singularities (see examples below).

As the condition $v=0$ is satisfied on the free surface, then its evolution is described by the parametric equation $z=ut+F(u)$ or, in the real form, by the equations
\begin{equation} \label{Eq_16}
x=ut+\mbox{Re}\,F(u),\quad \quad
y=\mbox{Im}\,F(u),
\end{equation}
where the horizontal velocity $u$ plays the role of a parameter (it lies in the interval $-V<u<V$). Together with the equations for the solid walls, $x=\pm Vt$, this equation defines the flow domain. It should be noted that the function $F$ can be presented as a formal infinite series
\[
F(U)= \textrm{i} \sum _{n=0}^{\infty }h_{n} \exp \left(\frac{ \textrm{i} \pi n(U+V)}{2V} \right)
\]
with arbitrary real coefficients $h_{n}$, i.e., in the form of a $4V$-periodic function.

\section{Examples of flows}

As an example, let us consider the flow described by \eqref{Eq_9} with the function $F$ defined by a comparatively simple expression:
\begin{equation} \label{Eq_17}
F(U)=\frac{ia}{b-\exp\,(i\pi U/V)}
\end{equation}
($a$ and $b$ are real constants). The flow described by \eqref{Eq_9} and \eqref{Eq_17} is symmetric with respect to the straight line $x=0$. At $a<0$ and $b<-1$, the free surface disturbance defined by  \eqref{Eq_17} is directed downward (see Figs.~2 and 3). The walls push the waves ahead of them; when the wave collide, these waves either capture a bubble (this situation occurs at $b^{*}<b<-1$, where $b^{*} =-2-\sqrt{3} \approx -3.73$) or form a cuspidal point (at $b\le b^{*} $).

\begin{figure}
\begin{center}
\resizebox{0.65\textwidth}{!}{\includegraphics{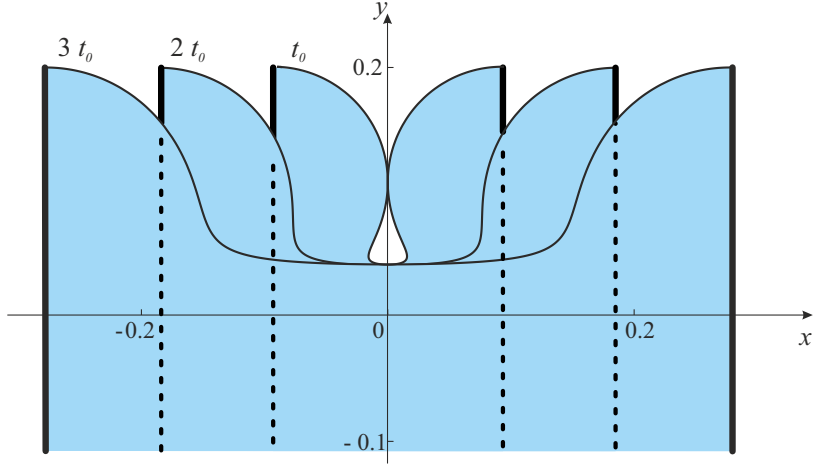}}
\end{center}
\caption{ Formation of a bubble on the free boundary. The flow is described by  \eqref{Eq_9} and \eqref{Eq_17} with $a=-0.1$, $b=-1.5$, and $V=2$. Consecutive time instants are shown: $t=3t_{0} ,\; 2t_{0} ,\;$ and $t_{0} $, where $t_{0} \approx -0.0463$ is the solution destruction instant. The vertical rays correspond to the solid walls.}
\label{disturb}
\end{figure}

The bubble formation dynamics is illustrated in Fig.~2 for $a=-0.1$, $b=-1.5$, and $V=2$. The flow is destroyed at the time $t_{0} \approx -0.0463$ owing to the fact that the boundary crosses itself (two free surfaces collide at one point). Correspondingly, the solution exists at $t\le t_{0} $.

The dynamics of the formation of a cuspidal point on the fluid boundary is shown in Fig.~3 for $a=-0.1$, $b=-4$, and $V=2$. It is formed when the singular point from the domain outside the flow reaches the free surface at the time $t^{*} =\pi a/(V(1-b)^{2} )$. In this case, the problem solution exists for a limited time $t\le t^{*} <0$.

\begin{figure}
\begin{center}
\resizebox{0.65\textwidth}{!}{\includegraphics{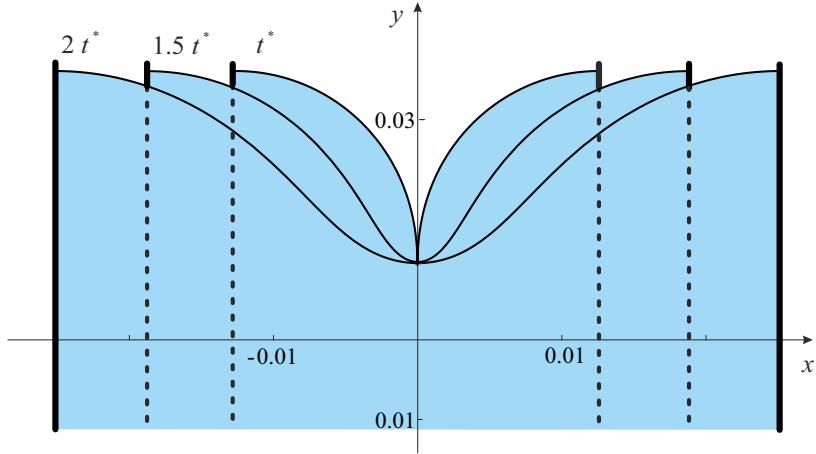}}
\end{center}
\caption{Formation of a cuspidal point in a fluid with a free surface. The flow is described by  \eqref{Eq_9} and \eqref{Eq_17} with $a=-0.1$, $b=-4$, and $V=2$. Consecutive time instants are presented: $t=2t^{*} ,\; 1.5t^{*} ,\;$ and $t^{*} $, where $t^{*} \approx -0.0063$ is the solution destruction instant.  }
\label{pic3}
\end{figure}
The free surface disturbance is directed upward at $a>0$ and $b>1$ (see Fig.~4). In this case, there is a tendency to droplet formation owing to fluid displacement induced by approaching of the solid walls. The solution exists for a limited time $t\le t^{**}$, where $t^{**} =-\pi a/(V(1+b)^{2} )$ is the instant of solution destruction caused by singularities reaching the boundary of the domain occupied by the fluid (in this case, the velocity field becomes multi-valued). Two scenarios are observed depending on the value of the parameter $b$. At $1<b<b^{**}$, where $b^{**} =2+\sqrt{3} \approx 3.73$, the droplet being formed has not enough space between the walls; for the solution to be continued up to the time $t^{**} $, the walls have to be bounded from above at the height $y=a/(1+b)$. Then the solution resembles squeezing of ice-cream from a wafer briquette (see Fig.~4, where $a=0.1$, $b=1.5$, and $V=2$). If $b\ge b^{**} $, the droplet has enough space to be accommodated between the vertical walls until the solution destruction instant $t^{**}$.

\begin{figure}
\begin{center}
\resizebox{0.85\textwidth}{!}{\includegraphics{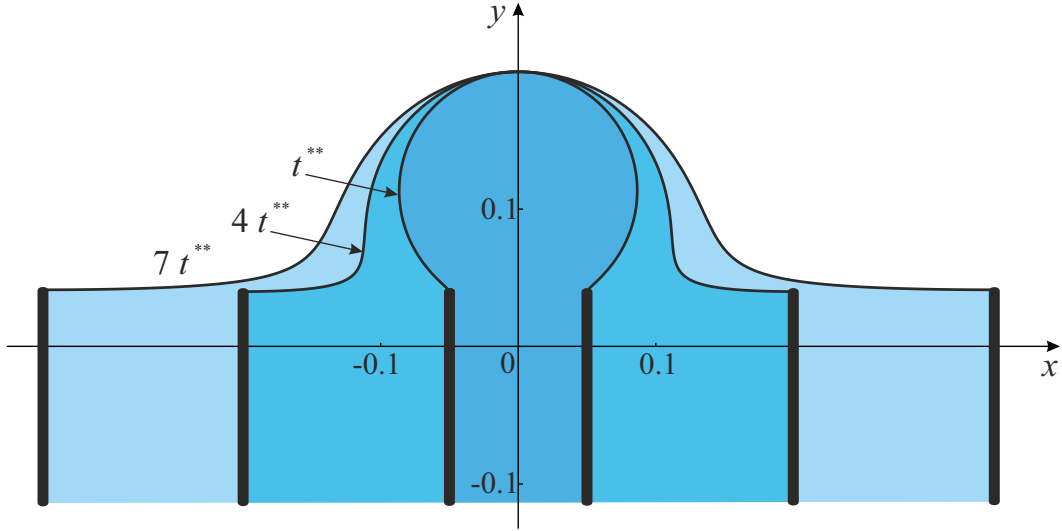}}
\end{center}
\caption{Formation of a droplet. The flow is described by \eqref{Eq_9} and \eqref{Eq_17} with $a=0.1$, $b=1.5$, and $V=2$. Three domains marked with different blue halftones corresponding to the fluid at three different time instants $t=7t^{**} ,\; 4t^{**}, \;$ and $t^{**} $ are presented. Here $t^{**} \approx -0.025$ is the solution destruction instant. }
\label{pic4}
\end{figure}
Figure~5 demonstrates the final (i.e., those that refer to the solution destruction instant $t^{**}$) shapes of the fluid boundary for different values of $b$. Here we take $a=b^{2}-1$, which ensures an unchanged (equal to two) difference of the fluid height on its free boundary (difference between the ordinates of the highest point of the free boundary and the droplet neck) for the parameter $b$ being varied. As $b$ decreases, the tendency to droplet formation becomes more pronounced. The width of the droplet neck $d$ is determined on the basis of the wall positions at the time $t^{**}$, i.e., it is described by the formula $d=-2Vt^{**} =2\pi a/(1+b)^{2}$. In the limit as $b\to 1$, a circular droplet with a unit radius is obtained for $a=b^{2}-1$, and the width of its neck tends to zero due to the relation $d\approx \pi (b-1)$.

\begin{figure}
\begin{center}
\resizebox{0.7\textwidth}{!}{\includegraphics{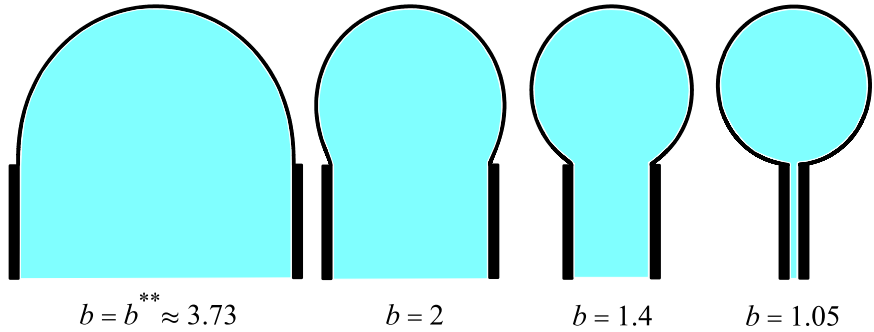}}
\end{center}
\caption{Final (i.e., those that refer to the solution destruction instant $t^{**} $) shapes of the domain occupied by the fluid for different values of $b$, $a=b^{2}-1$, and $V=2$. As $b$ decreases, a tendency to the formation of a circular droplet is observed.}
\label{pic5}
\end{figure}
Thus, even by using a comparatively simple expression \eqref{Eq_17} for the arbitrary function $F$ involved into the exact solution \eqref{Eq_9}, we obtain sufficiently complicated and interesting unsteady flows of the fluid with a free surface, leading to the formation of various singularities (droplets, bubbles, and cuspidal points) within a finite time.

\section{Conclusions}

It is clear from the overall considerations that the general solution of problem \eqref{Eq_1}--\eqref{Eq_6} has to contain a pair of arbitrary functions: one function is responsible for the initial shape of the free boundary, and the other one is responsible for the initial distribution of velocities. The solution \eqref{Eq_9} found in the present study  is a particular solution because it contains only one arbitrary function $F$. For this reason, the initial shape of the free boundary may be defined arbitrarily, but the velocity field is not arbitrary: it is defined by relation \eqref{Eq_9}. A specific feature of the class of solutions found in this study is the absence of the vertical component of velocity on the free boundary, which significantly restricts the flow types described by our approach. However, the use of even one arbitrary function can be considered as a noticeable achievement. As was shown by several examples, our solutions are able to describe some important processes, such as the formation of bubbles, droplets, and cuspidal points. The question about the possibility of complete integration of \eqref{Eq_1}--\eqref{Eq_6} remains open. Possibly, the general solution can be derived  as an expansion of the function $f(U,t)$ (see presentation \eqref{Eq_8dop}) in powers of $t$. For the fluid not bounded by the walls, the possibility of using an iterative procedure for constructing the general solution was discussed by Zakharov (2020). The approach in that study (Zakharov 2020) was based on conformal mapping of the domain occupied by the fluid onto a half-plane (Dyachenko \textit{et al.} 1996, Dyachenko \textit{et al.} 2021), which is principally different from the hodograph transformation applied in our investigation. It should be noted that integrability is additionally evidenced by the presence of new constants of motion, which arise during integration around singularities located outside the fluid (Dyachenko \textit{et al.}2019). The validity of the results (Dyachenko \textit{et al.}2019) can be verified by using the solutions obtained in the present study.

As was noted above, a significant drawback of flows with an unbounded free surface considered earlier by Karabut \& Zhuravleva (2014), Zubarev \& Karabut (2018), Zakharov (2020) (they are also described by solution (9)) was a non-typical behavior at the periphery. For an undisturbed flow \eqref{Eq_7} without walls bounding the fluid, the horizontal component of velocity $u=x/t$ diverges both as $\left|x\right|\to \infty $ and as $t\to 0$. Because of such an ``exotic" behavior of the velocity field, it is difficult to use the resultant exact solutions for the description of real flows. In particular, the solutions that described the base flow disturbances \eqref{Eq_7} were considered in (Zubarev \& Karabut 2018) as local solutions, which are valid only in a certain vicinity of the singularity being formed and meaningless at the periphery, i.e., at large values of $\left|x\right|$. The problem of divergence is eliminated in the present study, where the fluid is bounded by walls approaching each other with a constant velocity $V$. The horizontal velocity is always finite: it belongs
 to the interval $-V<u<V$ defined by the motion of the walls. The new solutions obtained in the study have a clear physical meaning; in our opinion, they will be included into a rather limited list of illustrative examples of exactly solved hydrodynamic problems.

\section{Acknowledgement} The work was supported in part by the Russian Foundation for Basic Research under Grant Nos. 19-01-00096 and 19-08-00098.

\section{Declaration of interests} The authors report no conflict of interest.

\end{document}